# Cognitive control and mental workload in multitasking

Philippe Rauffet[1], Sorin Moga[2], Alexandre Kostenko[1]

[1]Université Bretagne Sud, Lab-STICC UMR CNRS 6285

[2]IMT-Atlantique, Lab-STICC UMR CNRS 6285

## Synopsis

This study examines the relationship between mental workload and the cognitive control implemented in multitasking activity. A MATB-II experiment was conducted to simulate different conditions of multitasking demand, and to collect the behavioral and physiological activities of 17 participants. The results show that implementation of different modes of cognitive control can be detected with physiological indicators, and that cognitive control could be seen as a moderator of the effect of mental stress (task demand) upon mental strain (physiological responses).

## Background

Multitasking is a universal behavior allowing the management of simultaneous tasks. However, it can sometimes lead to performance decrement and mental overload. Mental workload is defined (ISO10075-1, 2018) with two dimensions: mental stress (task demand imposed upon operators) and mental strain (cognitive cost for operators). This workload is regulated by compensatory mechanisms (Hockey, 2003, Kostenko et al., 2019, Rauffet et al., 2016), allowing operators' engagement (Dehais et al., 2018) to be moderated towards greater performance (mental strain increases to meet task demand) or, conversely, towards mental stress relaxation (reducing strain and making it bearable for operators).

These mechanisms are encapsulated in the notion of cognitive control, allowing information processing to vary adaptively, depending on current goals, rather than remaining rigid (Braver, 2012). To characterize that, Hollnagel (1993) proposed a typology of 4 modes of cognitive control (MCC), from the most proactive to the most reactive one:

- **Strategic mode** is used when there is much time available and involves the management of simultaneous objectives, by adapting or generating new plans to control the situation.
- **Tactical mode** is based on the use of known rules to control a limited number of objectives.
- **Opportunistic mode** is implemented when available time is just enough. Operators focus on one single objective, determined by the most salient information.
- **Scrambled mode** occurs when the time available is extremely limited. The choice of action is random, the situation is no longer controlled.

This paper deeper investigates the articulation between cognitive control and mental workload. Considering MCC would allow to finer understand variations in mental workload, and help explaining certain discrepancies between subjective, physiological or performance measurement of mental workload (Hancock and Matthews, 2019).

## Methods

17 participants were recruited (whose 3 women, age = 21-45). Following works studying cognitive control in multitasking activity (Cegarra et al., 2017, Rauffet et al., 2020), MATB-II microworld was used.

This environment proposes to perform 4 tasks:
- **TRACK:** Keeping a position in a target using a joystick,
- **SYSM:** Detection of system alarms,
- **COMM:** Radio communications,
- **RESMAN:** Management of fuel resource levels in reservoirs.

The scenario is divided into four conditions of multitasking difficulty, each lasting 10 minutes. Varying the number of simultaneous tasks aims to generate different mental workload levels and different MCC (Figure 1).

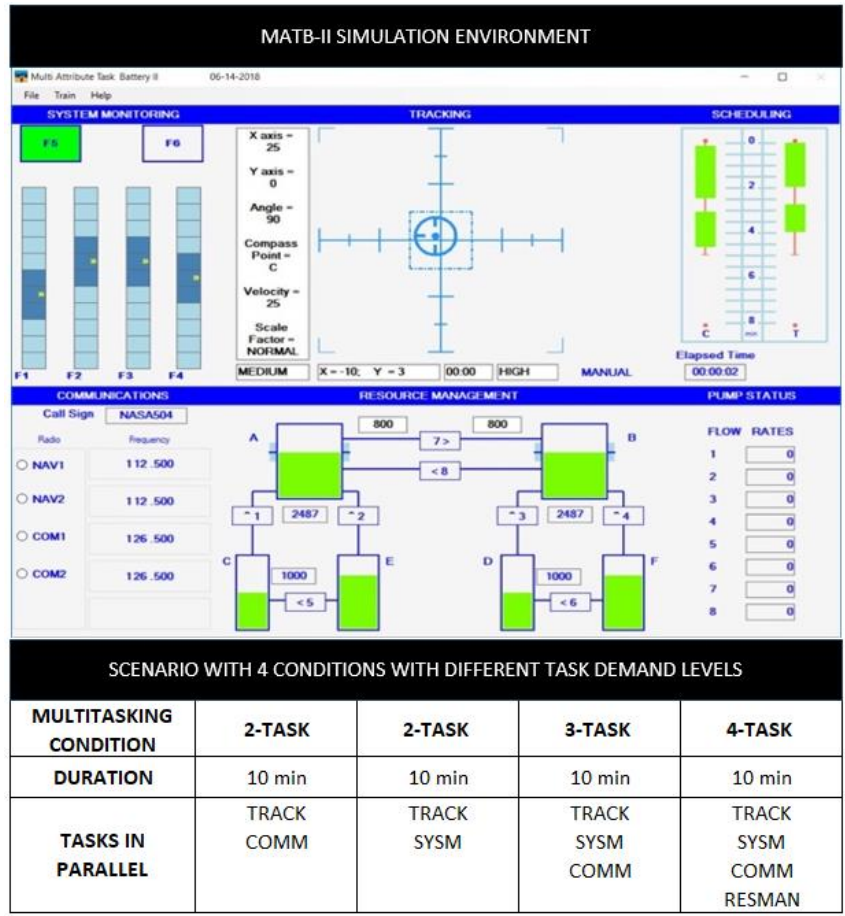

Figure 1: MATB-II Environment and experimental conditions.

Control modes were coded with participants' behavior on TRACK (presented as the most important task), using 3 criteria (time on target, time off target, joystick movements), over a fixed time window (30s). Tactical mode corresponds to good control (small adjustment movements, associated with an absence of position away from the target), scrambled mode to poor control (a position far from the target exceeding 5 seconds, combined with large joystick movements), and opportunistic mode to the presence of several medium movements or periods spent away from the target not exceeding 5 seconds. No strategic mode was observed.

Finally, data were collected to assess mental strain (with ocular and cardiac indicators acquired by using SeeingMachine© FaceLAB5 eyetracker and Zephyr© Bioharness belt) and MCC implementation (with indicators based on joystick interactions and MATB-II performance logs).

# Results

<u>Effect of mental stress (multitasking difficulty) on MCC implementation.</u> A two-factor Chi2 analysis revealed MCC significantly varied with multitasking difficulty ($X^2(6)=143.7$, $p<.001$). In 2-task conditions, TRACK+COMM resulted in mainly adopting opportunistic mode (48%; scrambled 17%), whereas TRACK+SYSM tended more towards tactical mode (54%; scrambled 10%). In 3-task condition, opportunistic mode was mostly adopted (49%; scrambled 25%). Finally, scrambled mode became the majority (49%, opportunistic 41%) in 4-task condition.

<u>Interaction effect between mental stress and MCC on mental strain (physiological responses).</u> lme4 R package (Bates, Maechler & Bolker, 2012) was used to perform linear mixed effects analyses of the relationship between multitasking difficulty and MCC (viewed as independent variables) and the neurophysiological indicators presented in Table 1 (dependent variables).

**Significant effects of multitasking difficulty were found on cardiac and ocular activities:**
- HRV was lower in 3- and 4-task conditions (greater mental strain) compared with the reference condition TRACK+COMM,
- Saccadic rate and pupil diameter increased with the number of tasks, whereas fixation duration decreased (higher visual effort).

**Significant effects of MCC were also observed, in interaction with multitasking difficulty:**
- Tactical and opportunistic modes led to a reduction in pupillary diameter compared to scrambled mode (especially in TRACK+COMM and TRACK+SYSM+COMM conditions),
- Tactical mode generates an increase in fixation duration compared to scrambled mode (especially in TRACK+COMM and TRACK+SYSM conditions).

|  |  | HRV<br>~ Multitasking<br>+ (1\|Participant) | Pupillary diameter<br>~ Multitasking * MCC<br>+ (1\|Participant) | Saccadic rate<br>~ Multitasking<br>+ (1\|Participant) | Fixation duration<br>~ Multitasking * MCC<br>+ (1\|Participant) |
|---|---|---|---|---|---|
|  | Intercept | 55.54 *** (3.20) | 3.76 *** (0.11) | 68.11 *** (6.34) | 0.31 *** (0.04) |
| **Multitasking Condition** | *Reference: 2-task TRACK+COMM* | - | - | - | - |
|  | 2-task TRACK+SYSM | -6.80 *** (0.52) | -0.11 *** (0.03) | 7.17 ** (2.55) | -0.06 * (0.03) |
|  | 3-task TRACK+SYSM+COMM | -3.99 *** (0.52) | 0.24 *** (0.02) | 14.78 *** (2.56) | -0.05 * (0.03) |
|  | 4-task TRACK+SYSM+COMM+RESMAN | -4.22 *** (0.66) | 0.44 *** (0.02) | 35.63 *** (3.23) | -0.12 *** (0.02) |
| **MCC** | *Reference: Scrambled mode* |  | - |  | - |
|  | Opportunistic mode |  | -0.07 *** (0.02) |  | 0.02 (0.02) |
|  | Tactical mode |  | -0.09 *** (0.02) |  | 0.07 *** (0.02) |
| **Multitasking * MCC** | *Reference: TRACK+COMM * Scrambled mode* |  | - |  | - |
|  | TRACK+SYSM * Opportunistic |  | 0.11 ** (0.03) |  | 0.00 (0.03) |
|  | TRACK+SYSM+COMM * Opportunistic |  | 0.03 (0.03) |  | -0.04 (0.02) |
|  | TRACK+SYSM+COMM+RESMAN * Opportunistic |  | 0.01 (0.03) |  | -0.01 (0.03) |
|  | TRACK+SYSM * Tactical |  | 0.10 ** (0.03) |  | -0.04 (0.03) |
|  | TRACK+SYSM+COMM * Tactical |  | 0.02 (0.03) |  | -0.08 ** (0.03) |
|  | TRACK+SYSM+COMM+RESMAN * Tactical |  | 0.12 ** (0.04) |  | -0.09 * (0.04) |
|  | AIC | 7156.83 | -1126.73 | 10405.37 | -1488.55 |
|  | Log Likelihood | -3572.31 | 577.36 | -5196.69 | 758.27 |
|  | Num. obs. | 1076 | 1074 | 1063 | 1076 |
|  | Num. groups: Participant | 14 | 14 | 14 | 14 |
|  | Var: Participant (Intercept) | 141.44 | 0.17 | 516.46 | 0.01 |
|  | Var: Residual | 42.06 | 0.02 | 1002.48 | 0.01 |

*** $p < .0001$, ** $p < .01$, * $p < .05$

*Table 1: Estimates of fixed effects from linear mixed-effect models, fitted by minimizing AIC function, for physiological and ocular responses (HRV means Heart Rate Variability). Contrasts are shown in relation to reference values.*

## Discussion

This study allows to articulate cognitive control with mental workload.

<u>Relationship between MCC and mental stress.</u> For 3- and 4-task conditions (the most difficult ones), we observed an increase in the adoption of scrambled mode (25% and 49% compared to less than 10% in TRACKING+COMM condition), associated with an increase in mental effort and less focused attention (saccades, duration of fixations), due both to an increase in visuo-spatial search, and to difficulty in maintaining priority on TRACK task.

<u>Relationship between MCC and mental strain.</u> Tactical mode required less effort than scrambled mode (significant effect on pupil diameter and duration of fixations). This is in line with the works of Hollnagel (1993) and Cegarra et al (2017), showing a link between cognitive control and mental stain, where scrambled control may result in more physiological activation than tactical control.

<u>"Moderating" role of MCC in stress-strain relationship.</u> Cognitive control could be seen as a moderator of stress-strain relationship: at an "equal" level of task demand, we could analyze finer physiological variations, due to operator regulations. We could therefore distinguish between variations that are 'normal' (the busier the activity, the greater the strain to meet demand), and smaller variations due to MCC.

## Disclosures

This work was supported by DGA, Thales AVS and Dassault Aviation.